# Seismic performance of an infilled moment-resisting steel frame during the 2016 Central Italy Earthquake

Phan Hoang Nam[1,2,*], Fabrizio Paolacci[1], Phuong Hoa Hoang[2]
[1]Department of Engineering, Roma Tre University, Rome, Italy
[2]Faculty of Road and Bridge Engineering, The University of Danang – University of Science and Technology, Da Nang, Vietnam
[*]Corresponding author, E-mail: phnam@dut.udn.vn

**Abstract:** A sequence of earthquakes occurred between the end of August 2016 and the end of October 2016 in Central Italy causing significant damage and major disruption in a wide area. The sequence of events is composed of five events with magnitude between Mw 5.5 to 6.5. As a consequence, numerous residential buildings in the affected area was not particularly resistant to the shaking, resulting in the collapse and heavy damage. With a particular focus on masonry infilled steel frames, this paper evaluates the seismic performance of an infilled moment-resisting steel frame located in Amatrice, Central Italy, which suffered significant damage during the August 2016 Central Italy earthquake. The aim is to investigate the effect of the masonry infill to the seismic performance of the building. The three-dimensional (3D) frame building is modeled using the Opensees software, where the beam and column elements are modeled by using a nonlinear hinge model and the infill is idealized as diagonal struts with nonlinear hysteretic behavior. Nonlinear static and dynamic analyses are performed for both bare and infill frames in order to assess the effect of the masonry infill on the overall seismic response and confirm the actual damage pattern surveyed in the aftermath of the 2016 Central Italy earthquake of the case study.

## 1. INTRODUCTION

The 2016-2017 Central Italy earthquakes consisted of several moderately-high magnitude earthquakes between moment magnitudes Mw 5.5 and 6.5. The first main shock on August 24th, 2016 was of moment magnitude Mw 6.0 and occurred at a depth of 4 km. The epicenter of the August's shock was located in the North-East area of the province of Rieti; this is a zone of high seismic hazard that has seen damaging earthquakes in the recent past. A series of aftershocks followed the above event until the Mw 5.9 Ussita earthquake that occurred on October 26th and the main-shock of October 30th (Mw 6.5 and depth of 9 km). The latter was the strongest event since 1980. The particular event had an epicenter 5 km from Norcia, overlaying the northern part of the fault activated by the earthquake of August 24th. These events caused a total of 299 fatalities, 386 injured and about 4800 homeless (Fiorentino et al. 2018). Most of the victims were in the areas of Amatrice, Accumoli, and Arquata del Tronto. In these municipalities, heavy damage and collapse of residential buildings were reported (De Risi et al. 2018, Fiorentino et al. 2018).

The observation of the damage has been primarily reported in existing steel multi-story residential buildings which exhibited low energy absorption and inadequate dissipation capacity. The insufficient horizontal stiffness of the frame and the masonry infill led to significant lateral drifts and buckling in the steel components, especially in the columns. Local damage (e.g., buckling) has also been observed at the connections due to the strut-action induced by the presence of the masonry infill.

The issues of the existing steel moment resisting frames (MRFs) are first related to the accurate prediction of the building seismic performance that is due to the complex behavior and the interaction between beam-to-column connections, as well as between the frame and masonry infill (Di Sarno et al. 2017). Because of such complexity and the absence of a realistic, yet simple, analytical model, the effect of masonry infill panels is often neglected in the nonlinear analysis of building structures. This assumption may lead to the inadequate prediction of the lateral stiffness, strength, and ductility of the structures. It also leads to the uneconomical design of the frame since the strength and stiffness demand on the frame could be largely reduced (El-Dakhakhni et al. 2003).

With a particular focus on masonry infilled steel frames, the aim of this paper is to investigate the seismic behavior of a steel MRF building with masonry infill panels. The building is two-story and located in Amatrice, Central Italy, which suffered significant damage during the 2016 earthquake. The three-dimensional model of the building is implemented using Opensees software, where both material and geometry nonlinearities are taken into account for. The infill is modeled as an equivalent diagonal strut, where its behavior is followed by a hysteretic model of axial force and displacement. The static pushover analysis is first performed for both bare and infilled models, and the nonlinear time history analysis is later performed with the Amatrice earthquake signal. The analysis results demonstrate the necessity of considering the infill in the steel frame model.

## 2. DESCRIPTION OF CASE STUDY

The sample steel frame is a two-story building located in Amatrice, Central Italy. It was built in the early 90's according to the 1996 Italian seismic code and consists of a basement, the ground floor, and two upper stories alongside a shorter top story that serves as a penthouse. The building plan layout is trapezoidal; it is 22.5-m-long and 6.6-8.45-m-wide. The ground floor height is 3.72 m and the interstory height of the first and second floor is 3.2 m. The flooring systems consist of concrete slabs on a steel corrugated 10 mm thick sheet. The steel grade for all structural components of the frame is equivalent to S235 ($f_y$ = 235 MPa). The columns are HEA 200, while the beams are HEA 300 and HEA 160. The beam-to-column connections are fully welded (i.e., moment-resisting frame). The infills consist of a double layer of perforated brick of the size of $12 \times 25 \times 8$ mm. The roof is supported by a truss structure, whose elements are made of 2L60x6 mm profiles.

Due to the lack of existing design drawings, the self-weight of the slab is assumed to equal to 4.88 kN/m², while the live loads are assumed equal to 2.00 kN/m² for the intermediate floors. The snow and wind loads are taken equal to 3.76 kN/m² and 0.96 kN/m², respectively.

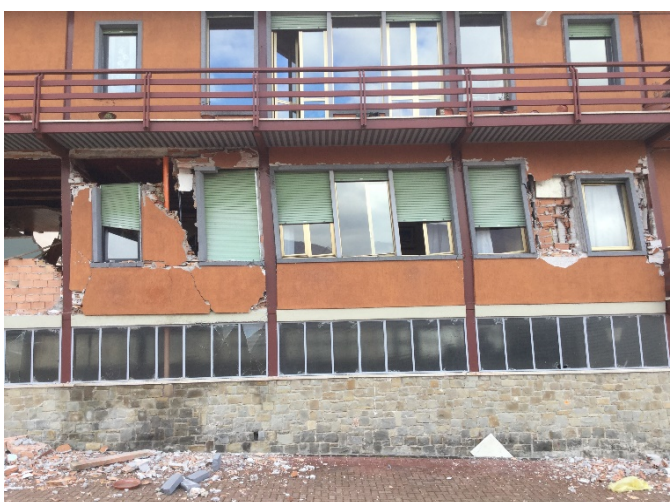

Figure 1. The sample building after the 2016-2017 Central Italy earthquakes

After the 24 August event, the building mainly suffered major cracks in the brick infill panels, with only small local flange instabilities observed at the top of two front columns of the ground floor. At the end of the entire seismic sequence, the building experienced evidently permanent deformation along its longer direction. Such permanent deformation was localized at the second level of the building with a visible residual interstory drift, as shown in Fig. 1. Preliminary finite element analyses of the building confirmed that the fundamental period of the structure is approximately equal to 0.56 s (Di Sarno et al. 2017). This was an uncoupled translational mode along the long side, which was mainly attributed to the orientation of the steel columns with their strong axes aligned with the short side of the building. Naturally, residual drift developed along the longitudinal axis.

## 3. NUMERICAL MODELING OF SAMPLE BUILDING

### 3.1 Infill modeling

In the past decades, numerous researches were conducted to investigate the behavior of steel frames infilled by masonry walls. Basically, two main approaches were proposed by various authors for modeling the masonry infill. The first is based on a full finite element (FE) modeling and the other one that is more simplified is based on the concept of "equivalent strut".

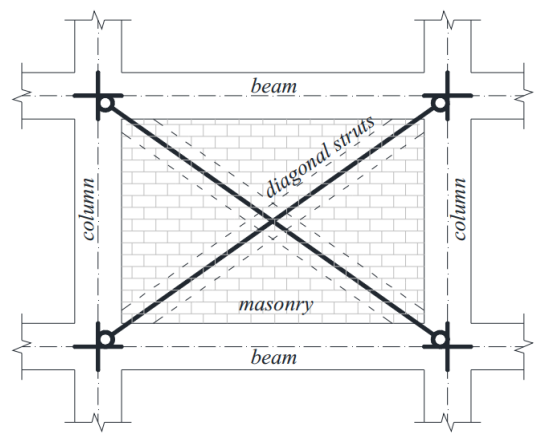

Figure 2. Macro-modelling of masonry infill through equivalent struts

Since FE models are generally unfit for practice-oriented analyses of the seismic response of steel frames, several macro-models were proposed and are currently available in the scientific literature to simulate the global seismic response of masonry infilled steel frames (Asteris et al. 2011). Such models are generally based on the assumption of two diagonal equivalent struts connecting the two opposite corners of the structural frame cell, as shown in Fig. 2. Although in principle, such macro-modeling approaches generally result in effective simplifications, the calibration of equivalent struts is not an easy task. In the following, a nonlinear force-displacement relationship to describe the response of equivalent struts proposed by Panagiotakos and Fardis (1996) is presented.

The skeleton curve is step-wise linear in shape to describe the compression-only behavior of the strut. In particular, the model is characterized by the following stress states (see Fig. 3):
- initial elastic behavior;
- post-elastic linear response characterized by a reduced value of stiffness;
- softening response of the panel after the maximum force;
- residual axial strength after a given value of displacement.

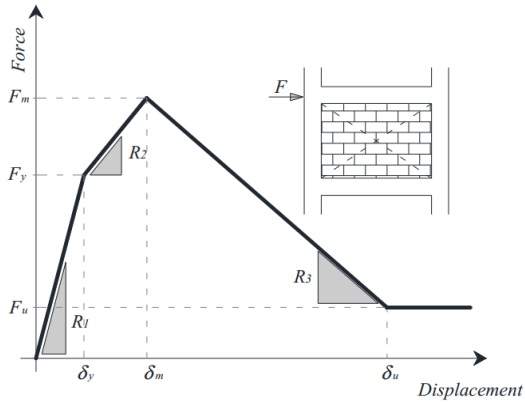

Figure 3. Force-displacement curve according to the model by Panagiotakos and Fardis (1996)

The initial shear stiffness $R_1$ of the uncracked panel can be defined as follows:

$$R_1 = G_w t_w l_w / h_w \quad (1)$$

where $G_w$ is the shear modulus of the masonry infill obtained in diagonal-compression tests while $t_w$, $l_w$ and $h_w$ are the thickness, the length and the height of the masonry wall. Then, the load value at the onset of cracking, namely yielding force $F_y$, is defined as follows:

$$F_y = f_{ws} t_w l_w \quad (2)$$

in which $f_{ws}$ is the tensile strength evaluated by diagonal-compression tests.

The axial post-cracking stiffness $R_2$ can be defined according to the following equation:

$$R_2 = E_w t_w w / d \quad (3)$$

in which $E_w$ is the Young modulus of masonry and the following equation can be adopted for evaluating the width w of the equivalent strut:

$$w = 0.175(\lambda h_w)^{-0.4} d \quad (4)$$

where $\lambda$ and $d$ are defined as:

$$\lambda = \sqrt[4]{\frac{E_w t_w \sin 2\theta}{4 E_c I_c h_w}} \quad (5)$$

$$d = \sqrt{h_w^2 + l_w^2} \quad (6)$$

in which $\theta$ is the slope angle of the infill's diagonal, $E_c$ is the elastic modulus of concrete, and $I_c$ is the moment of inertia of the column.

The post-cracking branch keep growing up to the maximum force $F_m = 1.3 F_y$. Then, a post-peak softening branch follows, whose (negative) stiffness $R_3$ can be assumed within the range $0.005 R_1 < R_3 < 0.1 R_1$ and the residual strength $F_u$ can range between $0 < F_u < 0.1 F_y$.

### 3.2 Structural modeling

The structural system is modeled with a refined three-dimensional finite element model using the finite element software Opensees (Mazzoni et al. 2007). The rigid diaphragm is assumed at the first and second floors to account for the presence of the concrete slab on the steel corrugated sheet. It is noted that the roof does not have any diaphragm. The building is assumed fixed at the base and the beam-to-column connections are assumed rigid and full strength (i.e., moment resisting).

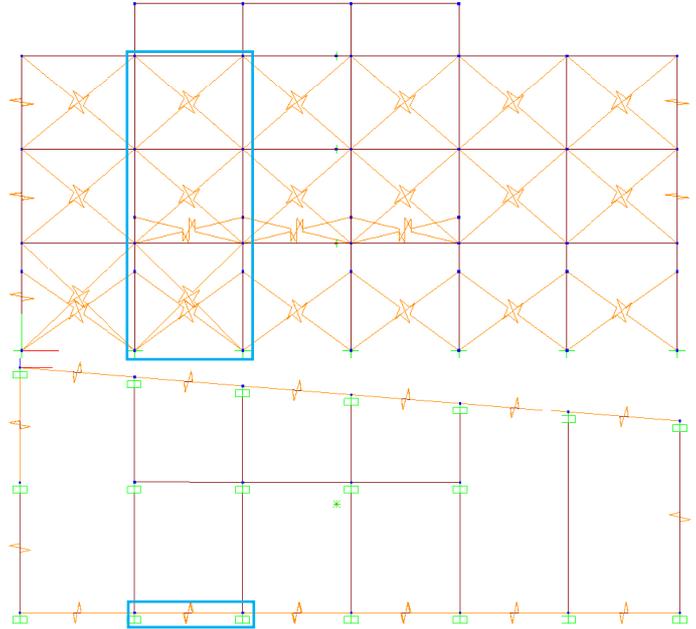

Figure 4. Floor plan and front elevation of the building model

The beams and columns are modeled using fiber sections with a behavior represented by the Menegotto-Pinto constitutive law. Two node-link elements are employed to model equivalent struts, which represent the nonlinear behavior of the masonry infill in terms of both strength and stiffness. The stiffness degradation in both loading and unloading phases is implemented through the so-called Pinching 4 material model available in Opensees. In detail, the floating point values defining force and deformation points on the skeleton compressive response are evaluated from the geometric and mechanical properties of the masonry infills through the model by Panagiotakos and Fardis (1996), while the ones on the response envelope in tension are assumed equal to zero. The floor plan and front elevation views of the model are shown in Fig. 4.

## 4. SEISMIC RESPONSE ANALYSIS

### 4.1 Nonlinear static pushover analysis

The presence of large residual damage in the surveyed structure indicates the high level of seismic action to which the building was subjected during the Central Italian earthquake.

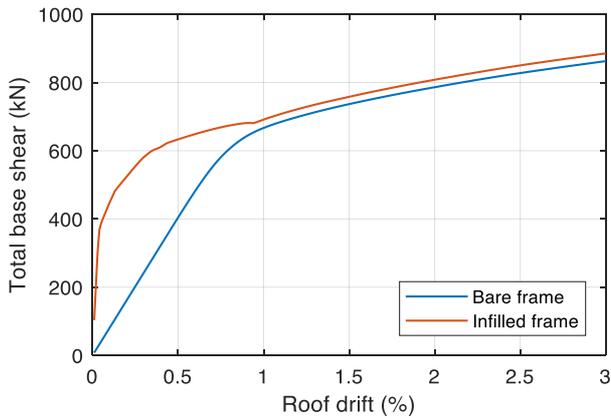

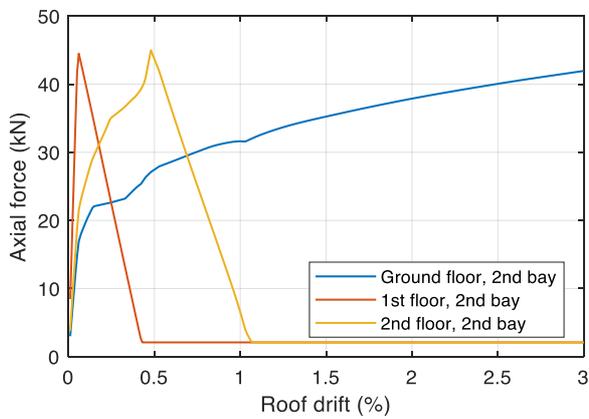

Figure 5. Static pushover analysis results of bare and infilled frame: (a) Roof drift - total base shear and (b) Roof drift - infill axial force

Consequently, nonlinear analyses of the building are conducted by using the model described in Section 3. As the first level of investigation, nonlinear static pushover analysis is conducted to assess the static behavior of the building with an increasing level of the action. The results in terms of the total base shear and the roof drift ratio of the building exhibit the significant increase both the stiffness and strength of the infilled frame as compared with those of the bare frame, as shown in Fig. 5(a).

Figure 5(b) shows a representative result of the pushover for the infill panels on three floors at the second bay (see Fig.4 marked blue color). The infill panel on the first floor is completely failed at a roof drift of 0.5%, while a much larger number of the roof drift, i.e., 1.0%, that can cause a total collapse of the one on the second floor. From this point forward, the pushover curves of the two cases are almost the same. The slight difference is due to the residual strength set for the infill to avoid the instability of the model. The external wall of the ground floor is implemented by the reinforced concrete; there is no damage observed. The results of the pushover analysis are in good agreement with the real damage of the building, where the significant damage observed was the masonry infill on the first floor.

### 4.2 Nonlinear time history analysis

The dynamic behavior of the building is investigated considering the East-West component of the ground motion recorded at the accelerometric station of Amatrice, as shown in Fig. 6(a), during the event M/S1, applied along the longitudinal direction. The *PGA* is equal to 0.85g with a resonance peak at 0.25 sec.

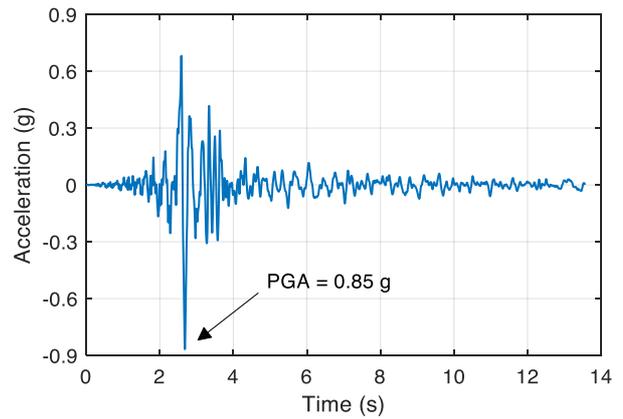

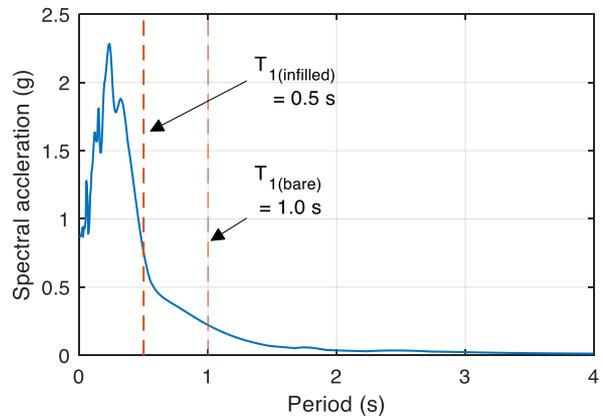

Figure 6. East-West record (a) and its elastic response spectrum (b) of the 24 August mainshock

The modal analysis is first carried out for both cases, with and without infill panels. The results show that the

ID No./ pp. …

fundamental period of the building with the infill decreases two times, $T_1 = 0.5$ s, as compared with that of the bare frame, $T_1 = 1.0$ s; this confirms the influence of the infill in increasing the overall stiffness of the building. The 5% damped response spectrum of the ground motion against the fundamental periods of the bared and infilled frames is shown in Fig. 6(b); this provides insight into the amplification of the spectral acceleration at $T_1 = 0.5$ s.

The nonlinear time history analysis is then carried out with the Amatrice earthquake. Figure 7 shows the time history of the total base shear for the bare and infilled frames. The response of the building evaluated with the time history analysis with respect to the weak direction, i.e., longitudinal direction, of the building. A significant contribution of the infill panels is observed. The peak base shear values for the bare and infilled frames are 800 kN and 1150 kN, respectively. The peak drift ratio along the building height is shown in Fig. 8, where a large value of the drift is observed for the first floor; this fits the real observation, as shown in Fig. 1.

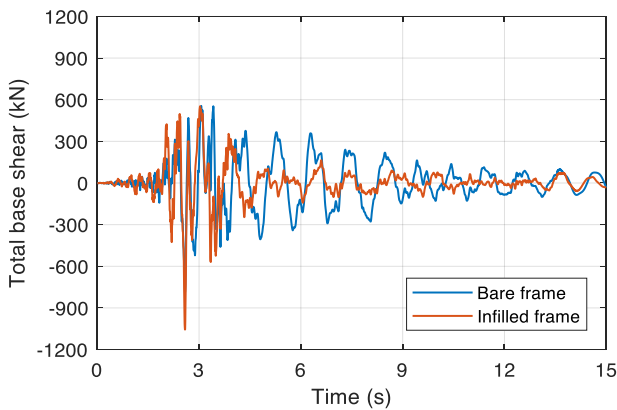

Figure 7. Time history data in terms of total base shear

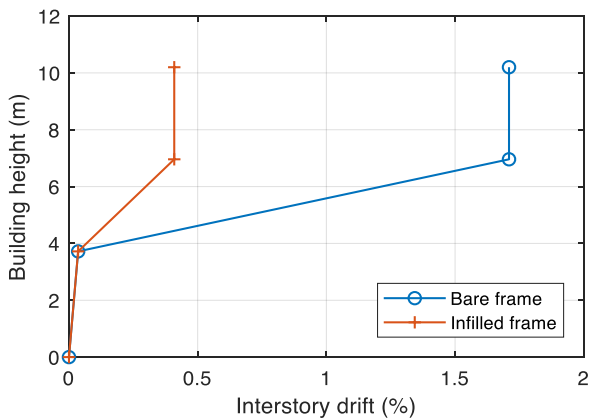

Figure 8. Peak interstory drift along with the building height

The local behavior of the infill panels on three floors is shown in Fig. 9. These three panels are in the second bay marked as blue color in Fig. 4. The most damaged infills are clearly located on the first floor of the building, where a large value of the interstory drift observed. The infill panel on the ground floor is still in the elastic range of the behavior, and the one on the third floor reaches the cracking strength. This result fits closely the observed response of the real building after the first event, i.e., the 2016 Amatrice earthquake, where most of the infill panels in the first floor suffered a major cracking. The damage in the steel elements is limited at first and second stories, which is similar to the real damage pattern observed on site.

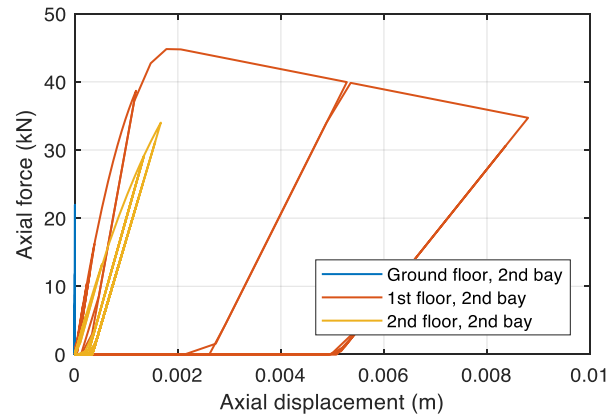

Figure 9. Axial force – displacement of the representative infill panels in the floors

## 4. CONCLUSIONS

The paper presents a numerical assessment of the structural performance of a case steel MRF building located in Amatrice, Central Italy, which suffered extensive structural and non-structural damage during the recent swarm of events. Refined numerical models of the existing steel building are developed with and without considering the effect of the perimeter masonry infill. The primary aim is to assess the effect of the masonry infill on the overall seismic response of the building and to confirm the actual damage pattern surveyed in the aftermath of the 2016 Central Italy earthquake for the case study.

The modal analysis results emphasize the significant effect of the masonry infill on the global dynamic behavior of the structure. It is found that the fundamental period of the vibration of the steel frame with the infill is nearly 1/2 of that of the bare steel structure, 0.5 seconds versus 1.0 seconds. Thus, it is of paramount importance, when estimating the seismic demand on steel framed buildings. The presence of the infill tends to augment the lateral stiffness of the structural system and, in turn, to increase the seismic loads.

The seismic performance of the case study is then assessed through advanced, comprehensive and efficient static and dynamic nonlinear analyses, which based on a 3D nonlinear model of the building. It is found that the analyses can provide a realistic damage observation when the masonry infill is included in the structural model and can thus be utilized to perform a reliable risk assessment of the building.

ID No./ pp. …